# Performance Evaluation of Transcriptomics Data Normalization for Survival Risk Prediction

Ai Ni[1], Li-Xuan Qin[2]

[1] Division of Biostatistics, College of Public Health, Ohio State University.

[2] Department of Epidemiology and Biostatistics, Memorial Sloan Kettering Cancer Center.

**ABSTRACT**

One pivotal feature of transcriptomics data is the unwanted variations caused by disparate experimental handling, known as handling effects. Various data normalization methods were developed to alleviate the adverse impact of handling effects in the setting of differential expression analysis. However, little research has been done to evaluate their performance in the setting of survival outcome prediction, an important analysis goal for transcriptomics data in biomedical research. Leveraging a unique pair of datasets for the same set of tumor samples – one with handling effects and the other without, we developed a benchmarking tool for conducting such an evaluation in microRNA microarrays. We applied this tool to evaluate the performance of three popular normalization methods – quantile normalization, median normalization, and variance stabilizing normalization – in survival prediction using various approaches for model building and designs for sample assignment. We showed that handling effects can have a strong impact on survival prediction, and that quantile normalization, a most popular method in current practice, tends to underperform median normalization and variance stabilizing normalization. We demonstrated with a small example the reason for quantile normalization's poor performance in this setting. Our finding highlights the importance of putting normalization evaluation in the context of the downstream analysis setting and the potential of improving the development of survival predictors by applying median normalization. We make available our benchmarking tool for performing such evaluation on additional normalization methods in connection with prediction modeling approaches.

## INTRODUCTION

Survival analysis plays a foundational role in cancer transcriptomics studies for developing reliable predictors of patient prognosis and treatment response[1-4]. While statistical methods are available to address the issues of high dimensionality and signal sparsity in these studies, research is still lacking on the issue of unwanted data variations associated with disparate experimental handling, which is a pivotal feature of transcriptomics data[5-8]. Many of these studies borrowed existing methods for data normalization that were developed in the setting of differential expression analysis for group comparison, when signals are the difference between group means[9-12]. While the performance of these normalization methods has been extensively studied in group comparison, little research has been done to re-evaluate their performance in the setting of survival analysis, when signals are associated with a censored outcome, partly due to lack of awareness and dearth of benchmarking tools[13-18].

We set out to develop the much-needed benchmarking tool and conduct such an assessment in microRNA microarrays[19,20]. Our approach leverages a pair of datasets for the same set of tumor samples that we previously collected. One dataset was collected with uniform handling to minimize handling effects; the other was collected without uniform handling and exhibited handling effects[21,22]. The uniformly-handled dataset enabled estimation of the biological effects for each sample, serving as 'virtual samples', and the difference between the two arrays for each sample allowed estimation of the handling effects for each array in the non-uniformly-handled dataset, serving as 'virtual arrays'. The virtual samples and the virtual arrays were then reassigned and rehybridized to generate additional data under various scenarios for evaluating normalization methods[21,23]. In addition, outcome data were simulated by sequentially reallocating the observed outcomes to the virtual samples with probability weighting to achieve a prespecified level of association[24].

In this paper, we report this benchmarking tool based on re-sampling and our findings on how survival prediction accuracy is impacted by the application of three popular normalization methods: quantile normalization[13], median normalization, and variance stabilizing normalization[25]. In addition to post-hoc data normalization, we also considered the use of two study designs for assigning arrays to samples: randomized and sorted by survival time (leading to longer survival times being assigned to earlier arrays and shorter survival times to later arrays).

## METHODS

### Collection of the Empirical Data

Ninety-six high-grade serous ovarian cancer samples and 96 endometroid endometrial cancer samples were collected at Memorial Sloan Kettering Cancer Center between 2000 and 2012. Their microRNA expression levels were measured using Agilent microarrays (Release 16.0, Agilent Technologies, Santa Clara, CA) twice, each with a different experimental design. In the first design, the 192 samples were handled by a single experienced technician in one experimental run and assigned to the arrays in a balanced manner via the use of randomization and blocking (each slide of eight arrays serves as an experimental 'block'). In the second design, the same samples were handled by two technicians in multiple batches over time and assigned to arrays in the order of sample collection. We call the first design the 'uniformly-handled' design and the second the 'non-uniformly-handled' design. Further details on data collection can be found in the articles by Qin et al.[21,22].

**Estimation of Biological Effects and Handling Effects from the Empirical Data**

*Estimation of biological effects*   We used the data collected with the uniformly-handled design for the 96 ovarian-cancer samples as a best approximate for the biological effects of these samples. We call them 'virtual samples'.

*Estimation of handling effects*   Assuming that handling effects were additive, we used the differences between the two arrays for each tumor sample (one from each design) to estimate the handling effects for each array in the non-uniformly-handled dataset. We call them 'virtual arrays' and split them (by whole slides) into two equal-sized sets, one set with the first 96 arrays for prognosticator training and another with the last 96 arrays for validation. The additivity assumption has been deemed reasonable for microarray data and has been adopted in published methods on microarray data normalization and analysis [26,27].

*Elicitation of regression coefficients*   We used the data for the 96 ovarian-cancer samples from the uniformly-handled design to assess each microRNA's association with progression free survival (PFS), an important survival outcome variable in ovarian cancer[28]. PFS is defined as the time from primary surgery to disease progression, death, or loss of follow-up, whichever occurs first. The rate of censoring was 23% (22/96) in our data. Univariate Cox regression analysis showed that six markers have p values less than 0.005 and regression coefficients ranging from 0.26 to 0.78 [29]. For the purpose of the simulation study, we generated three sets of 'true regression coefficients' from the estimated regression coefficients at three different signal levels to elucidate the impact of true marker effect size on survival prediction.

1. Moderate signal: We used the quadruple of the estimated regression coefficients for the six significant markers and zero for the other markers.
2. Weak signal: We randomly chose 30 markers and set their regression coefficients to 0.35 and those of other markers to zero. The L1 norm of the weak-signal coefficient vector equals that of the moderate-signal coefficient vector, so the two vectors have the same "total" effect that was distributed differently among markers – one on a few markers with large effects and the other on many markers with small effects.
3. Null signal: We used a null regression coefficient vector with all zeros as a negative control.

**Simulation of Survival Outcome and Array Data for Prognosticator Training and Validation**

*Simulation of survival outcome*   We simulated new PFSs for the 96 virtual samples based on the true regression coefficient vectors. To ensure a realistic marginal distribution for PFS without having to arbitrarily assume a parametric distribution, we developed a permutation-based procedure to simulate PFS exhibiting various levels of association with biological effects. This method was inspired by a similar permutation-based method proposed by Heller[24]. It first sorts the observed PFS times in an ascending manner regardless of the censoring status, and then sequentially pairs them with the virtual samples starting from the smallest time as follows.

i. At PFS time $t_{(j)}$, the probability of choosing virtual sample $i$ from those that have not been chosen is calculated as Pr(choose virtual sample $i$ at time $t_{(j)}$) $= p_{ij} = \frac{\exp(x_i \beta_0)}{\sum_{k \in R_j} \exp(x_k \beta_0)}$, where $x_i$ is the biological effect of virtual sample $i$, $\beta_0$ is the true regression coefficient vector, $R_j$ is the set of virtual samples that have not been chosen by $t_{(j)}$.
ii. Since $\sum_{i \in R_j} p_{ij} = 1$, the selection of virtual samples at $t_{(j)}$ is determined by a single realization of a multinomial distribution with $n = 1$ and $p_j = (p_{1j}, p_{2j}, \ldots, p_{k_j j})$, where $k_j$ is the size of $R_j$.
iii. The above steps are repeated through all sorted PFS times.

We conducted a small simulation to demonstrate that the above procedure leads to the intended association between the covariates and the survival outcome and present the results in the Appendix (S-Figure 4 and 5).

*Simulation of array data for prognosticator training*    Training microarray data were simulated using a process called 'virtual re-hybridization', so as to preserve the complex correlation structure of the biological effects and handling effects[21,23]. Namely, the 96 virtual samples (along with their PFS times) were reassigned to the 96 virtual arrays allocated for training, and handling effects for each virtual array were then added to the biological effects of the assigned virtual sample. We considered two scenarios for the reassignment: (1) randomized, and (2) sorted by simulated PFS time (leading to longer PFS times being assigned to arrays handled by one technician and shorter PFS times to arrays by the other) so that PFS time was associated with handling effects. To examine whether the magnitude of handling effects impacts the performance of the normalization methods, we also simulated training data with augmented handling effects by tripling the values in the virtual arrays before adding them to the virtual samples.

*Simulation of array data for prognosticator validation*    Data were simulated either with handling effects following the same virtual re-hybridization method, except that the 96 virtual arrays allocated for validation were used instead, or without handling effects (mimicking a scenario when the validation dataset was of better data quality).

*Simulation scenarios*    Table 1 summarizes the eight scenarios investigated in our simulations. The scenarios are arranged in such a way that handling effects are increasingly prevalent and involved (that is, associated with the survival outcome). Each scenario used 400 simulation runs.

**Prognosticator Training and Validation**

*Array data preprocessing*    We assessed three normalization methods: (1) quantile normalization, (2) median normalization, and (3) variance stabilizing normalization. The same normalization methods were applied to the training and test data, with the latter in a 'frozen' manner [30]. Quantile normalization was applied to the training data using the *normalize.quantiles* function in the R package *preprocessCore*; the quantiles derived from the training data were used to apply frozen quantile normalization to the test data. Similarly, when median normalization was used for the training data, frozen median normalization was applied to the test data. Variance stabilizing normalization was carried out using the *vsn2* function in the R package *vsn*. In addition to handling effects adjustment via normalization, the data were additionally preprocessed with log2 transformation and median summarization across replicate probes for each marker[31].

**Table 1.** Summary of simulation scenarios.

| Scenario Notation[a] | Handling effects in training data | Handling effects in test data | Handling effects association with outcome in training data | Handling effects association with outcome in test data |
|---|---|---|---|---|
| HE00Asso00 | No | No | No | No |
| HE10Asso00 | Yes | No | No | No |
| HE10Asso10 | Yes | No | Yes | No |
| HE11Asso00 | Yes | Yes | No | No |
| HE11Asso10 | Yes | Yes | Yes | No |
| HE11Asso01 | Yes | Yes | No | Yes |
| HE11Asso11 | Yes | Yes | Yes | Yes |
| HE11Asso1-1 | Yes | Yes | Yes | Yes (negatively) |

[a] 'HE' stands for Handling Effects. The first and second digits following 'HE' indicate presence ('1') *versus* absence ('0') of handling effects in training and test data, respectively. 'Asso' stands for association with survival outcome. The first and second digits following 'Asso' indicate the presence ('1'=positive, '-1'=negative) *versus* absence ('0') of association between handling effects and survival outcome in training and test data, respectively.

*Prognosticator training*    We applied two commonly used methods for variable selection to build a multivariate Cox proportional hazard model for PFS prediction: (1) univariate filtering using the per-marker p-values and (2) regularized Cox proportional hazard regression. To reduce computational burden and alleviate collinearity in model fitting, we pre-filtered the markers using two criteria: (1) high abundance (mean expression on the log2 scale among the 96 samples >=8), and (2) no strong inter-marker correlation (Pearson correlation coefficient <0.9). These criteria were applied to each simulated dataset, and the set of markers that passed the filtering varied across simulation runs. Typically, 100 to 200 markers remained and entered the model-fitting step.

- In the univariate filtering analysis, we assessed PFS association for each marker using a univariate Cox proportional hazards regression, and selected markers with a p value less than or equal to a given cutoff. The selected markers were then included in a multivariate Cox proportional hazards regression model. The p value cutoff was selected via a grid search from 0 to 0.01 by 0.0005. The value that minimized the AIC from the multivariate Cox model was selected as the optimal cutoff.
- In the regularized regression, we first used the univariate Cox regression to select [$n_0$/4] markers that had the largest partial likelihood, where $n_0$ is the number of events in the training data and [.] denotes the nearest integer, and then performed regularized Cox regression with the selected markers using one of two penalties: (1) the LASSO penalty and (2) the adaptive LASSO penalty[32, 33]. This two-step variable selection strategy has been extensively studied in high-dimensional data literature[34-36]. Six-fold cross-validation was used for selecting the tuning parameters of these penalties.
- As a reference, we also fitted a non-penalized multivariate Cox regression model using the true predictive markers (six markers for the moderate-signal model and 30 markers for the weak-signal model), referred to as the oracle method. Although the oracle model is not obtainable in practice, these results are nevertheless revelatory for assessing the impact of handling effects and the performance of data normalization with regard to prediction accuracy.

*Prognosticator validation*     An ideal approach for validating a prognostication model is to assess its predictive accuracy in an independent test dataset. In our study, test data shared the same biological effects with training data but differed in handling effects and the random sample-to-PFS pairing. Hence the test-data validation mainly reflected the effectiveness of data normalization and the robustness of the resulted model to handling effects. Harrell's C-index was used to measure prediction accuracy[37]. The mean, $2.5^{th}$, and $97.5^{th}$ percentile of the C-index among the 400 runs for each simulation scenario were reported.

To summarize, in this simulation study, we assessed the performance of three methods for data normalization in combination with four approaches for prognostic model building using data generated under eight scenarios of handling-effect pattern and three levels of survival signal strength.

**RESULTS**

**Moderate Signal**

We present the results of the oracle, LASSO-penalized, and univariate-filtering methods in the main text. In addition, we include in the Appendix the result of the adaptive LASSO-penalized method, which is very close to that of LASSO-penalized method (Supplementary Figure 1 and 2).

*Oracle method*

The simulation result, in terms of prediction accuracy measured by the test data C-index, of the oracle method is presented in Figure 1.A. Among the three normalization methods, median normalization was the obvious best performer in all eight simulation scenarios, while quantile normalization tended to be the worst (closely following variance stabilizing normalization). In the presence of handling effects, the C-index was between 0.72 and 0.82 for median normalization, between 0.68 and 0.75 for variance stabilizing normalization, and between 0.64 and 0.73 for quantile normalization, depending on the specific handling-effect pattern. As a reference, when handling effects were absent in both training and test data, the C-index was 0.92, 0.80, and 0.74 for these three methods, respectively. Across the eight scenarios, as handling effects became more prevalent and outcome-associated, their negative impact on prediction became stronger. The most influencing factor was the method of normalization, followed by the presence of handling effects and then by their level of outcome association (comparing HE10Asso00 with HE11Asso00 *versus* with HE10Asso10).

*Penalized regression method*

Figure 1.B shows the simulation result when the LASSO-penalized Cox regression was used to build the prediction model. The relative performance of the three normalization methods stayed similar to that for the oracle method. That is, median normalization was the best and quantile normalization the worst across the eight patterns of handling effects. Their C-index ranged from 0.74 to 0.83, 0.69 to 0.76, and 0.71 to 0.76 for median, quantile, and variance stabilizing normalization, respectively, in the presence of handling effects in training data and/or test data; it was 0.92, 0.81, and 0.84 for the three methods, respectively, in the absence of handling effects. Compared to the oracle method, the LASSO method slightly improved the prediction accuracy by up to 0.07 across the simulation scenarios. This is likely due to the fact that LASSO tends to select more predictor markers into the final model than the true model.

*Univariate filtering method*

Figure 1.C presents the results when the univariate filtering method was used to build the prediction model. The relative performance of the normalization methods was again consistent with the oracle method. In the presence of handling effects, the C-index ranged from 0.74 to 0.81, 0.64 to 0.74, and 0.70 to 0.76 for median, quantile, and variance stabilizing normalization, respectively. In the absence of handling effects, it was 0.93, 0.88, and 0.90, for the three methods, respectively. Compared to the LASSO method, the prediction accuracy for the univariate filtering method was slightly worse (by up to 0.05) and substantially more variable.

In addition to the above results, we also performed simulations when the magnitude of handling effects was tripled in the training and test data. As expected, the prediction performance of the normalization methods worsened across all scenarios. Nevertheless, the relative performance of these methods remained the same. We therefore did not include the results in the paper.

**Weak Signal**

The simulation results of the three prediction modeling methods under weak signals are presented in Figure 2. They are very similar to those under moderate signals, supporting the robustness of our findings in terms of the performance of normalization methods to the size of predictive signal for survival outcome.

**Null Signal**

We further examined the performance of data normalization under the null model. Noting that the oracle method is not available under the null model and the simulation results for the other two methods were very similar, we present the result for the LASSO method in Figure 3 and that for the univariate-filtering method in the Appendix (S-Figure 3). Under the null model, as expected, the prediction model did not offer any value beyond a random guess regardless of the choice of normalization method in the first six scenarios. However, in the seventh scenario when handling effects existed in both training and test data with positive outcome-association (HE11Asso11), normalization led to a small improvement in prediction with C-index slightly over 0.5; conversely in the eighth scenario when the direction of association between handling effects and the survival time was opposite (HE11Asso1-1), normalization slightly harmed prediction with C-index below 0.5, suggesting that the prediction model performed even worse than a random guess. The observations in the last two scenarios showcased that handling effects that confound with the survival time can either induce a false positive predictor or dampen a true positive one, depending on the direction of the confounding association.

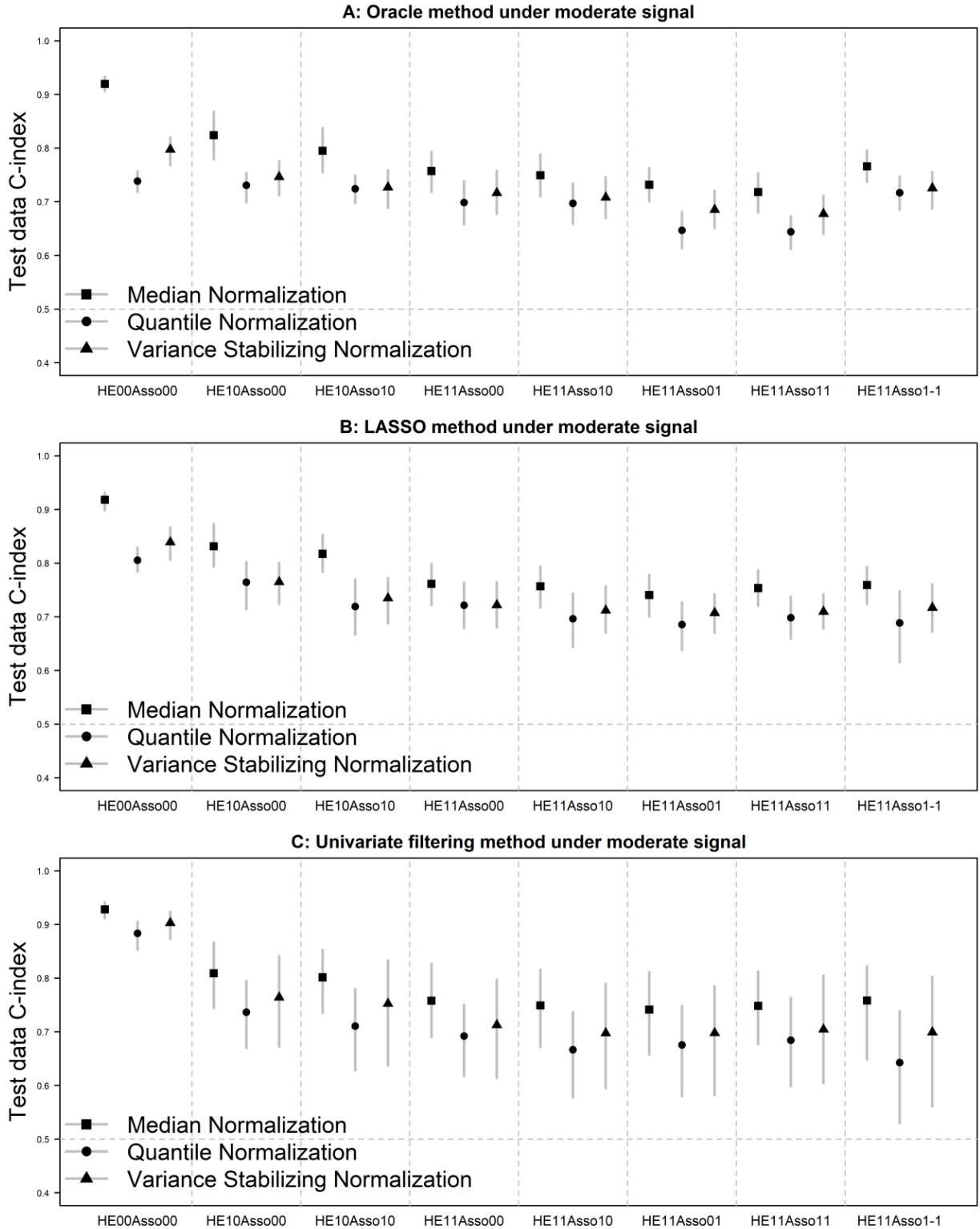

**Figure 1** Test data Harrell's C-index of the prediction model developed by the oracle, LASSO penalized, and univariate filtering methods under a moderate level of signal. Vertical bars represent 2.5[th] and 97.5[th] percentiles. Symbols in the bars represent mean values.

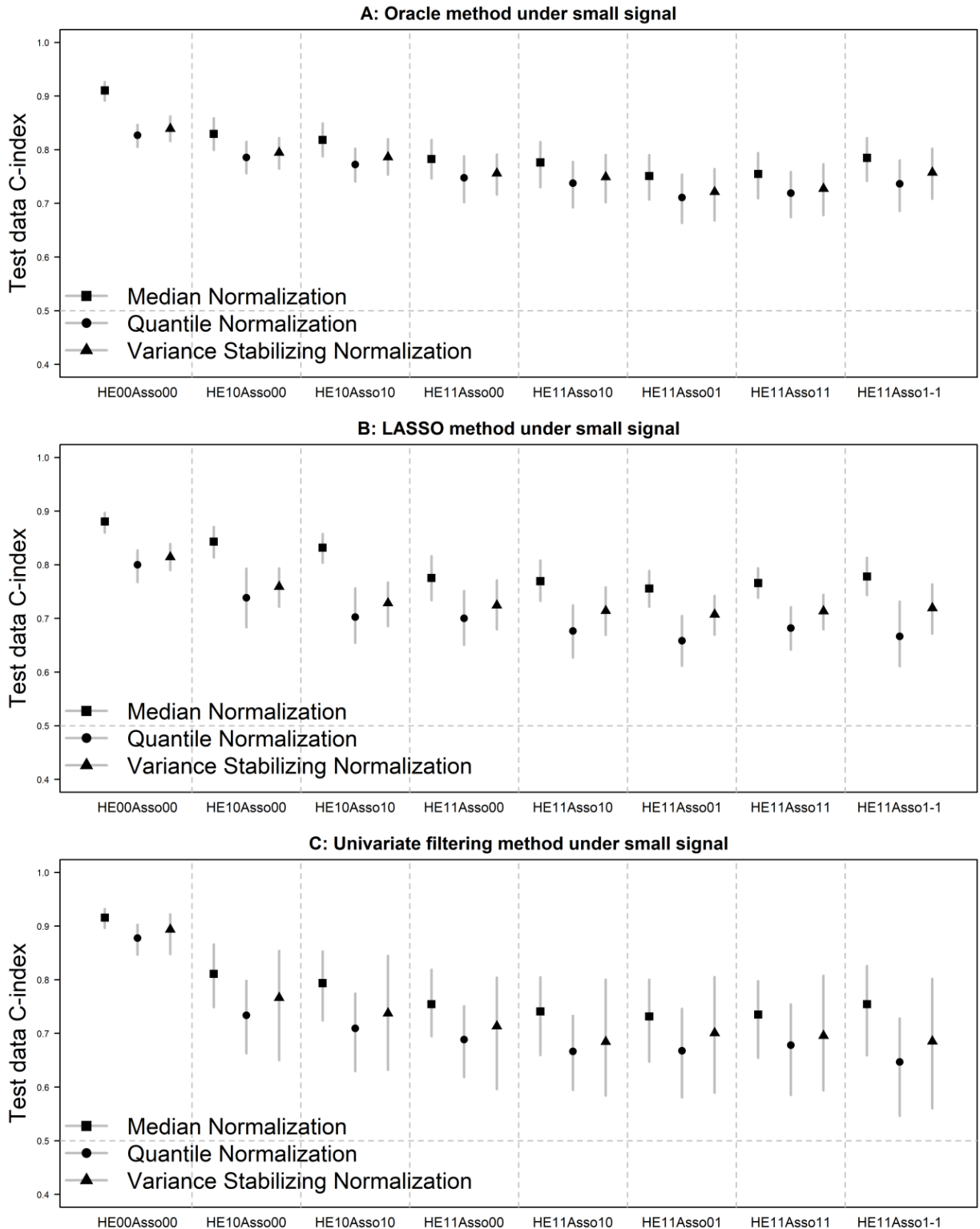

**Figure 2** Test data Harrell's C-index of the prediction model developed by the oracle, LASSO penalized, and univariate filtering method under a weak level of signal. Vertical bars represent 2.5[th] and 97.5[th] percentiles. Symbols in the bars represent mean values.

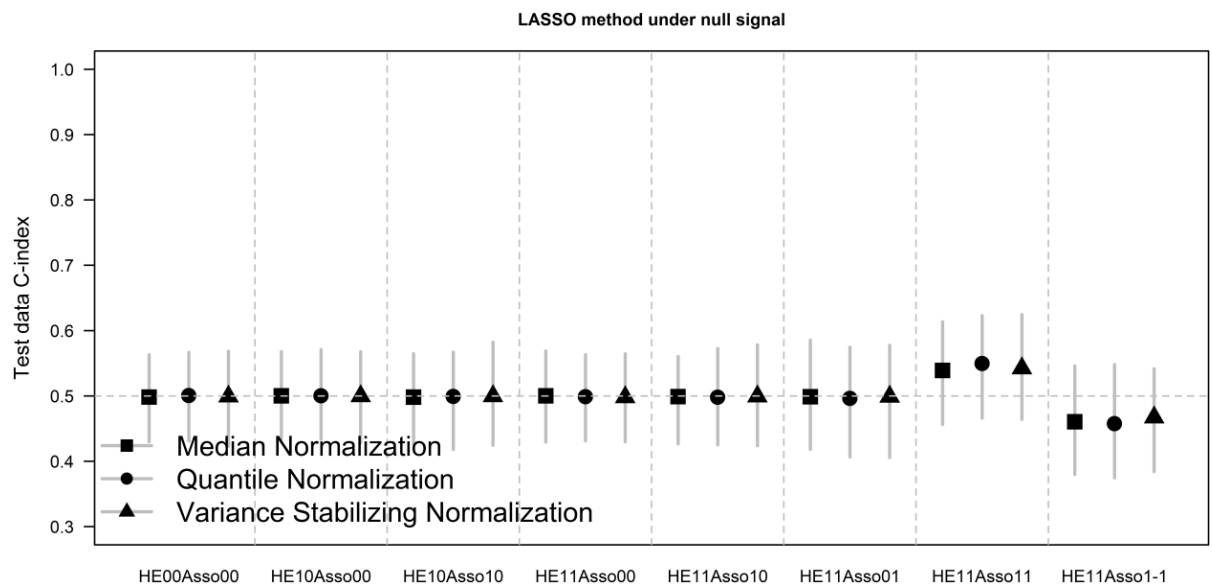

**Figure 3.** Test data Harrell's C-index of the prediction model developed by the LASSO penalized method under null signal. Vertical bars represent 2.5$^{th}$ and 97.5$^{th}$ percentiles. Symbols in the bars represent mean values.

**DISCUSSION**

Effective prognostic biomarkers of patient clinical outcomes are of keen interest in cancer research, as they can help identify high-risk population, tailor treatment options for patients, and design clinical trials for assessing new therapies. While proven useful for biomarker discovery and patient classification in the setting of group comparison, quantile normalization performed poorly for survival outcome prediction in our study. Intuitively, quantile normalization replaces the ranked data of a marker by the averaged ranked data across samples: for the purpose of biomarker discovery, it is effective for removing bias due to handling effects in the estimation of group mean difference; for the purpose of building a prediction model for time to event variables, however, it runs the risk of changing the rank of marker data across samples, which we show with numerical examples in the Appendix (S-Figure 4), and subsequently attenuating its regression coefficient towards zero, similar to the effect of adding noise to a predictor[38].

Median normalization, a runner-up in popularity to quantile normalization, performed better than quantile normalization for the purpose of survival outcome prediction. Its better performance may be explained by the fact that median normalization distorts the rank of marker data to a lesser extent as it only forces the median instead of all percentiles to be the same across samples (S-Figure 5).

The prediction performance of variance stabilizing normalization is only slightly better than quantile normalization. It involves a step where all markers of each sample are rescaled by a sample-specific affine-linear transformation[25]. This transformation may have resulted in rank distortion of marker data across samples similar to the effect of quantile normalization, hence the unsatisfactory performance in the setting of survival risk prediction.

Our simulation results revealed that penalized regression methods offer slightly more accurate and substantially less variable prediction than the univariate filtering method. This observation agrees with the general opinion in statistical literature that automated significance-based stepwise variable selection procedures are unstable, especially when the correlation among predictors is high[39]. In our study, adaptive LASSO method tended to give slightly sparser models than the LASSO method, but their prediction performances were similar.

Balanced sample assignment (via the use of study design elements such as blocking, randomization, and stratification) has been shown to be effective for avoiding the negative impact of handling effects when developing a predictor of a binary outcome[21]. For predicting time to event outcomes, blocking and stratification are no longer applicable; randomization (that is, random reassignment of virtual samples to virtual arrays) is still useful as shown in our study. Our study under null signals clearly demonstrated that, in the absence of randomization, spurious positive or negative predictive value (Figure 3 HE11Asso11 and HE11Asso1-1) could arise as a result of the association between survival outcome and handling effects.

To summarize, our study demonstrates the importance of evaluating the performance of normalization methods in the setting of survival prediction and provides a benchmarking tool for such evaluation for microRNA microarrays. Among the methods examined in this study, median normalization and penalized regression offer better survival risk prediction. We encourage interested researchers to use our tool for assessing additional methods for data normalization and prediction modeling that they use in their practice.

**DATA AVAILABILITY**

Human tumor tissues used in this study were obtained from participants who provided informed consent, and their use in our study was approved by the Memorial Sloan Kettering Cancer Center Institutional Review Board. The data and code supporting the conclusions of this article can be found at https://github.com/LXQin.

**FUNDING**

Supported by National Institutes of Health Grants No. CA214845 (L.X.Q.) and CA008748 (L.X.Q.).

# APPENDIX

**Supplementary Figures**

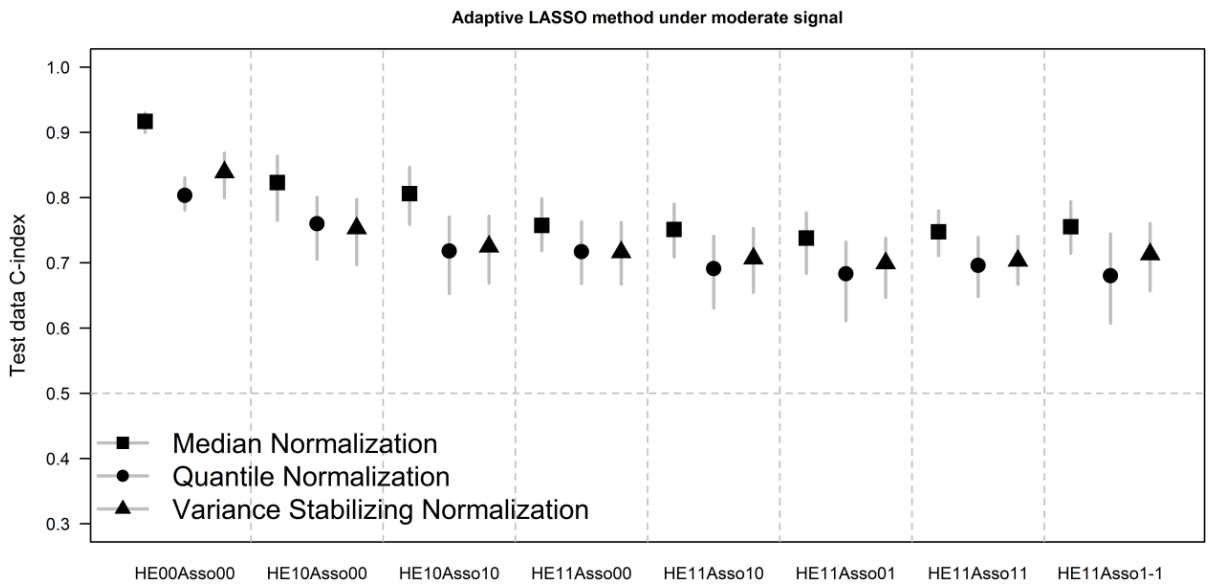

**S-Figure 1.** Test data Harrell's C-index of the prediction model developed by the adaptive LASSO penalized method under a moderate level of signal. Vertical bars represent 2.5[th] and 97.5[th] percentiles. Symbols in the bars represent mean values.

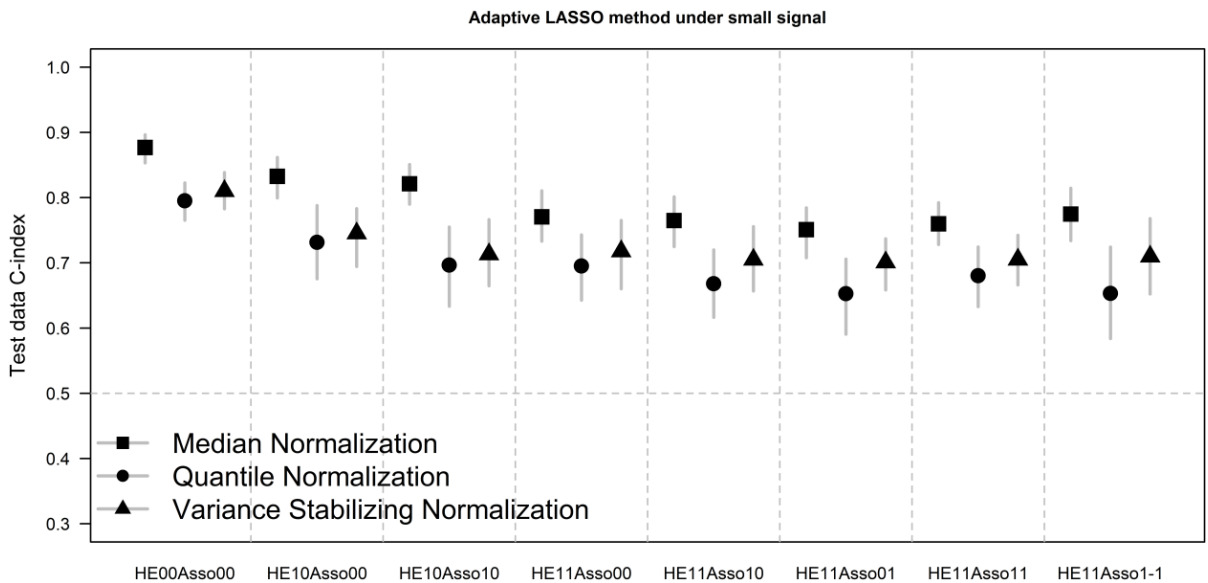

**S-Figure 2.** Test data Harrell's C-index of the prediction model developed by the adaptive LASSO penalized method under a weak level of signal. Vertical bars represent 2.5[th] and 97.5[th] percentiles. Symbols in the bars represent mean values.

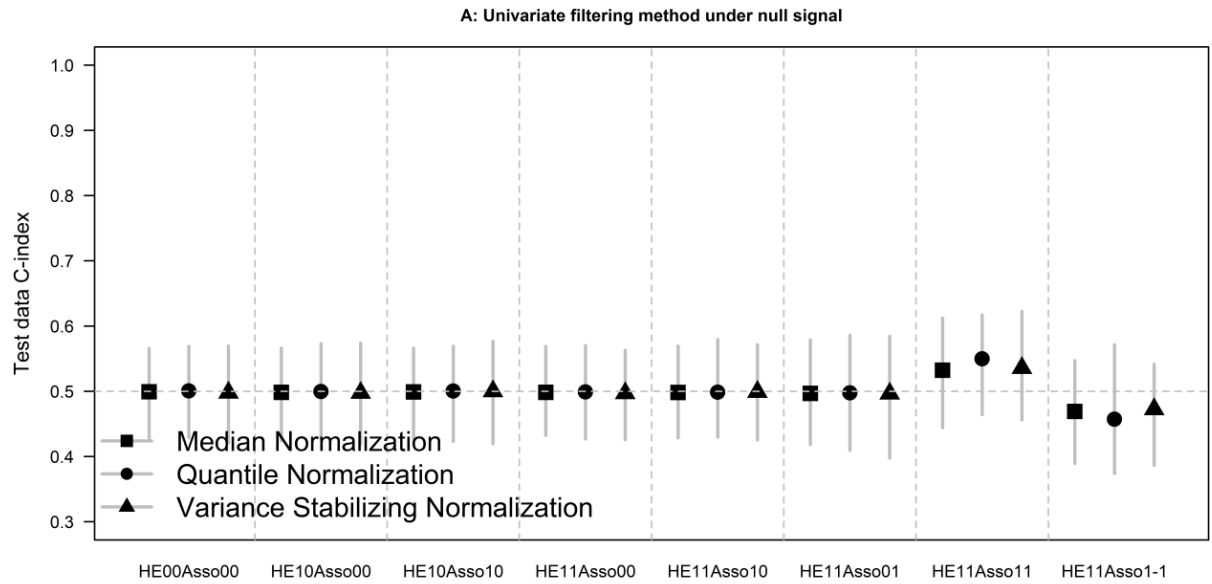

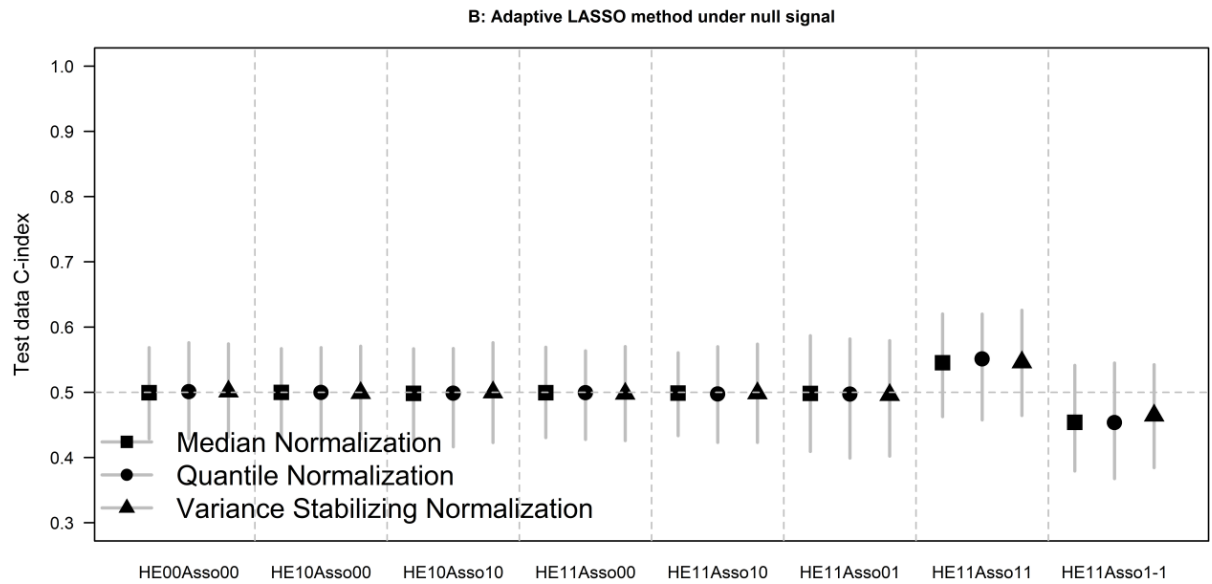

**S-Figure 3** Test data Harrell's C-index of the prediction model developed by the univariate filtering and adaptive LASSO penalized methods under null signal. Vertical bars represent 2.5[th] and 97.5[th] percentiles. Symbols in the bars represent mean values.

**Simulation on the algorithm for outcome-to-sample assignment**

We conducted a small simulation based on the real endometrial cancer dataset to demonstrate the performance of the proposed permutation-based procedure for survival outcome generation. In the simulation, we only included the six genes that are considered truly associated with the survival outcome plus one gene that is not associated with the survival outcome. The true regression coefficients were set to $b_0$=(1.04, 1.40, 1.45, 1.62, 3.13, 1.76, 0), where the first six components are the quadrupled effects of the six genes used in the simulations in the main text. The last component of value zero is for the additional gene unassociated with the survival outcome. The original PFS time and status in the endometrial cancer dataset were used in this simulation. Five hundred permutated datasets were generated using the proposed procedure and analyzed using Cox's proportional hazards model. The means of estimated regression coefficients are (1.10, 1.51, 1.53, 1.69, 3.29, 1.86, 0.05). The percent bias in the point estimates is mostly around 5%. Given the moderate sample size of the dataset, the bias is considered to be in a reasonable range.

**Detailed explanation and numerical examples on rank distortion by quantile and median normalization**

1. *Detailed explanation*.
   To simplify the argument we will consider data without handling effect. We first consider quantile normalization. Focus on any one of $p$ markers, say $X = (X_1, \ldots, X_n)$, where $n$ is the sample size. We denote its quantile normalized version as $X^{QN} = (X_1^{QN}, \ldots, X_n^{QN})$. Let $R_{mi}$ ($i = 1, \ldots, n$) denote the rank of $X_i$ in $X$ and $R_{si}$ denote the rank of $X_i$ among the $p$ markers in subject $i$. Similarly define $R_{mi}^{QN}$ and $R_{si}^{QN}$ for $X^{QN}$. Since $X_1^{QN}$ is the average of markers with rank $R_{si}$ across all subjects, we have $X_i^{QN} > X_j^{QN}$ whenever $R_{si} < R_{sj}$. However, $R_{si} < R_{sj}$ does not necessarily lead to $R_{mi} < R_{mj}$. Therefore, it is possible that $X_i < X_j$ when $X_i^{QN} > X_j^{QN}$, hence rank distortion in marker $X$ by quantile normalization. The same argument applies to all $p$ markers.

   Next we consider median normalization. Again we focus on any one of $p$ markers, say $X = (X_1, \ldots, X_n)$. Denote its median normalized version as $X^{MN} = (X_1^{MN}, \ldots, X_n^{MN})$. Let $M_i (i = 1, \ldots, n)$ be the median of the $p$ markers in subject $i$. Let $M$ be the value (usually mean of $M_i$ across $n$ subjects) that $M_i$ is normalized against. By the definition of median normalization, $X_i^{MN} = X_i - (M_i - M)$. It follows that, for any $i$ and $j$, $X_i^{MN} - X_j^{MN} = X_i - X_j - (M_i - M_j)$. Thus, when $X_i - X_j$ and $M_i - M_j$ have the same sign and $|X_i - X_j| < |M_i - M_j|$, the rank of $X_i$ and $X_j$ is reversed after median normalization. The same argument applies to all $p$ markers.

2. *A miniature numerical example.*
   Consider an artificial dataset with five subjects and three markers: $X_1 = (1,2,3,4,5), X_2 = (0,1,4,1,6), X_3 = (2,0,5,6,7)$. After quantile normalization, $X_1^{QN} = (3.2, 4.4, 1.8, 3.2, 1.8), X_2^{QN} = (1.8, 3.2, 3.2, 1.8, 3.2), X_3^{QN} = (4.4, 1.8, 4.4, 4.4, 4.4)$. None of the three markers retains its original rank after quantile normalization. After median normalization, $X_1^{MN} = (3.2, 4.2, 2.2, 3.2, 2.2), X_2^{MN} = (2.2, 3.2, 3.2, 0.2, 3.2), X_3^{MN} = (4.2, 2.2, 4.2, 5.2, 4.2)$. Again, none of the markers retains its original rank after median normalization.

3. *A real data example.*

    We used the expression data from the 96 ovarian-cancer samples from the uniformly-handled design (i.e. biological effect data) for this illustration. The original data was subject to quantile or median normalization. We plotted in S-Figure 4 and 5 the post-normalization rank against the original rank for the six markers with true nonzero regression coefficients (see section *Simulation of survival outcome* in the main text for details). If the original rank of a marker is completely reserved after normalization, the points in the plot should all lie on the diagonal line. The more scattered the points around the diagonal line, the worse the rank distortion is by normalization. It is obvious that both quantile and median normalization distort the original rank of markers, but median normalization exhibits much less distortion than does quantile normalization.

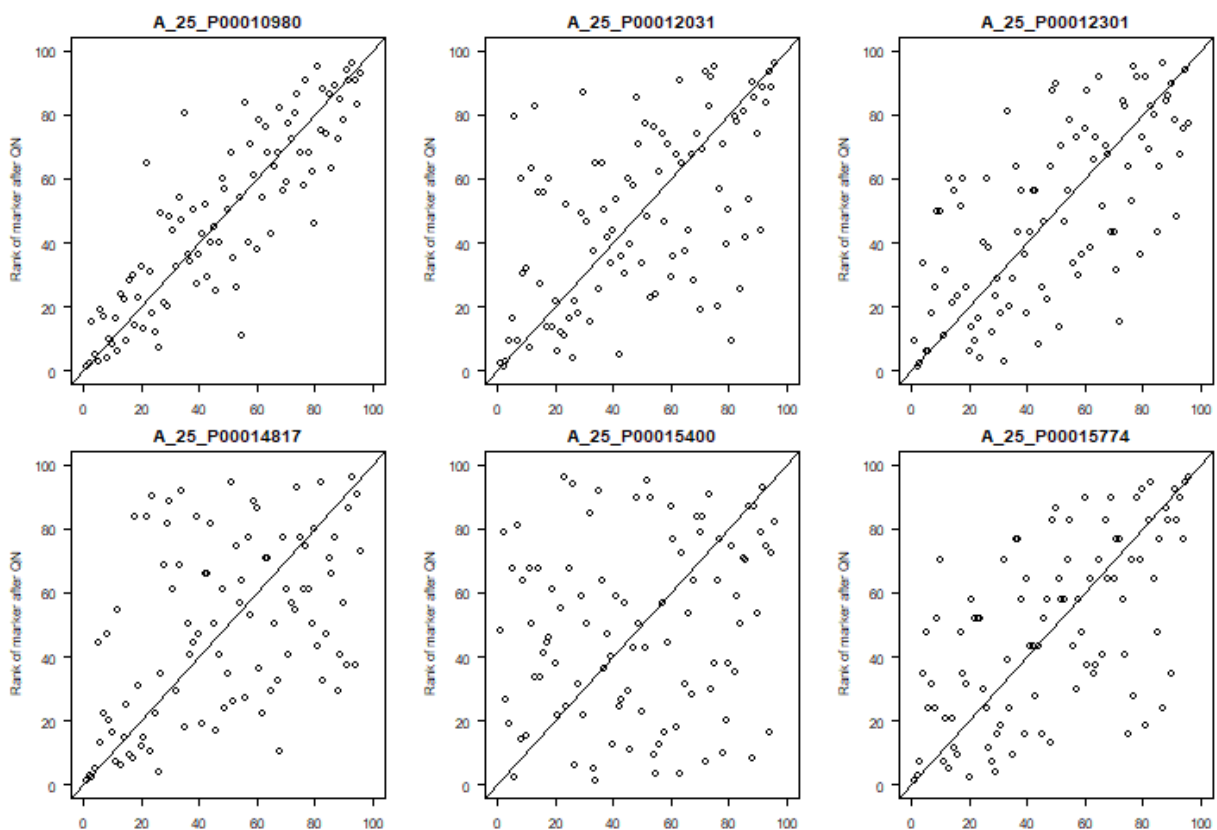

**S-Figure 4** Post-quantile-normalization rank vs original rank for the six markers with true non-zero regression coefficient. Each point represents one subject. The solid line represents the diagonal line. In all six plots, the points are widely scattered around the diagonal line, suggesting that quantile normalization substantially distorts the rank of markers.

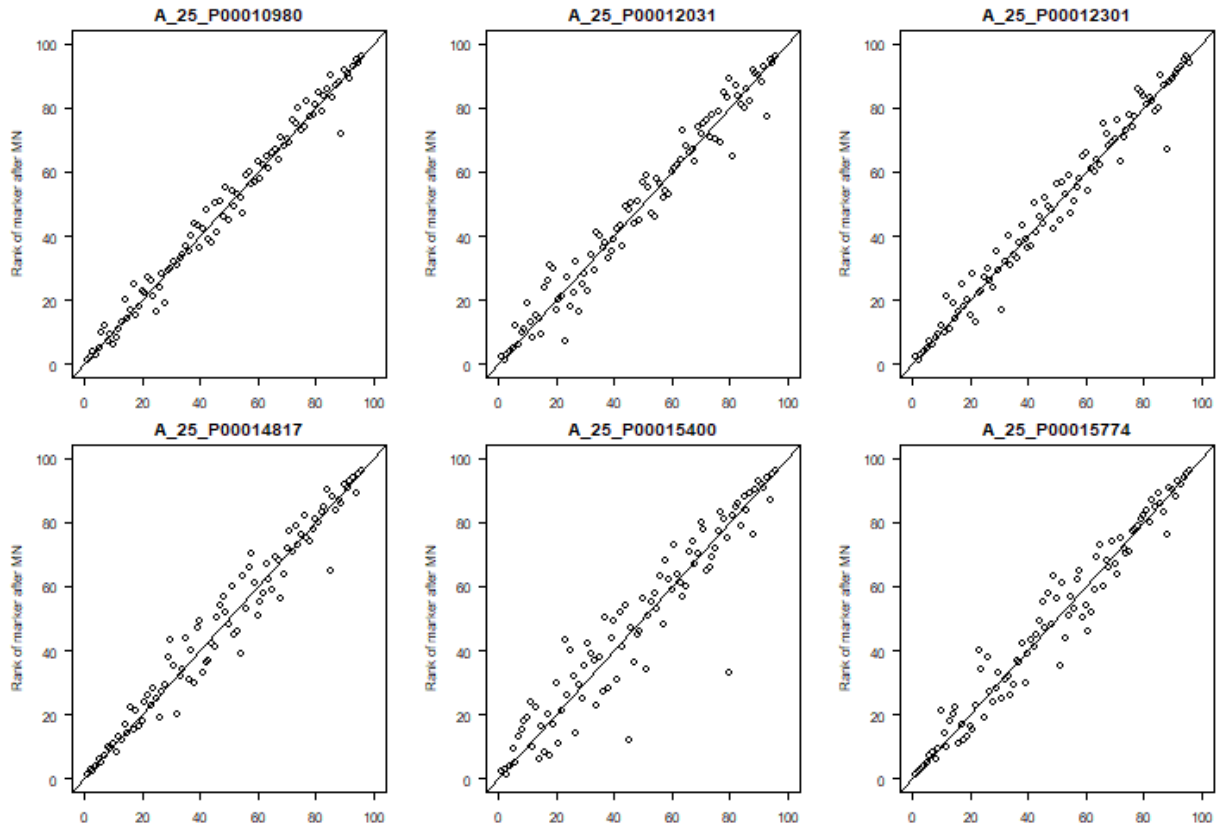

**S-Figure 5** Post-median-normalization rank vs original rank for the six markers with true non-zero regression coefficient. Each point represents one subject. The solid line represents the diagonal line. In all six plots, the points are scattered around the diagonal line, suggesting that median normalization distorts the rank of markers to a much less extent than quantile normalization as shown in S-Figure 4.

**R code to install the private R package under development that contains the data and code used in this study**

devtools::install_github("LXQin/precision.seq", auth_token = "5b5239aa86dfd6a45ecc069b44546f8c800274a4")